\documentclass[12pt]{iopart}

\usepackage{graphicx}

\begin{document}

\title[Charge noise in LISA]{Charge Induced Acceleration Noise in the LISA Gravitational Reference Sensor}

\author{Timothy J Sumner$^{1,2}$, Guido Mueller$^1$, John W Conklin$^3$, Peter J Wass$^3$ and Daniel Hollington$^2$}  

\address{$^1$Department of Physics, University of Florida, 2001 Museum Road, Gainesville, Florida 32603, USA}
\address{$^2$Imperial College London, Prince Consort Road, London. SW7 2AZ, UK}
\address{$^3$Department of Mechanical \& Aerospace Engineering, University of Florida, 231 MAE-A, P.O. Box 116250, Gainesville, Florida 32611, USA}
\ead{t.sumner@ufl.edu, t.sumner@imperial.ac.uk}
\vspace{10pt}
\begin{indented}
\item[]September 2019
\end{indented}

\begin{abstract}

The presence of free charge on isolated proof-masses, such as those within space-borne gravitational reference sensors, causes a number of spurious forces which will give rise to associated acceleration noise.  A complete discusssion of each charge induced force and its linear acceleration noise is presented. The resulting charge acceleration noise contributions to the LISA mission are evaluated using the LISA Pathfinder performance and design.  It is shown that one term is largely dominant but that a full budget should be maintained for LISA and future missions due to the large number of possible contributions and their dependence on different sensor parameters. 

\end{abstract}

%
%
%
%
%

\section{\label{intro}Introduction}

The build-up of charge on isolated free-flying proof-masses fully enclosed within spacecraft is inevitable due to cosmic-rays and solar energetic particles which are so highly penetrating that they can reach the enclosed proof-mass and interact with it and its surrounding environment~\cite{jafry96}. Once a proof-mass is charged there will be forces between it and its surrounding conducting enclosure which include an electrostatic force from its own mirror charges, an interplay with any applied or stray electric potentials and a Lorentz force from motion through ambient magnetic fields.   Associated with those forces will be induced acceleration noise which could be a limiting noise component for weak-force experiments, and the charge has then to be controlled~\cite{buchman95, armano18, sumner07, sumner04, sun06, pollack10}.  The level of control needed depends on the sensitivity of the sensor to charge. 

This study documents all the charge induced noise terms for gravitational reference sensors of the type used for LISA Pathfinder~\cite{armano16} and proposed for LISA~\cite{Amaro17}. In section~\ref{model} a sensor model geometry is defined and its charge sensitivity is derived.  In section~\ref{acc} the acceleration noise terms are identified and in section~\ref{budget} the overall noise budget is evaluated based on the LISA Pathfinder sensor performance.      
  
\section{\label{model}The Sensor Model}
\begin{figure}[h]
\includegraphics[width=0.8\textwidth]{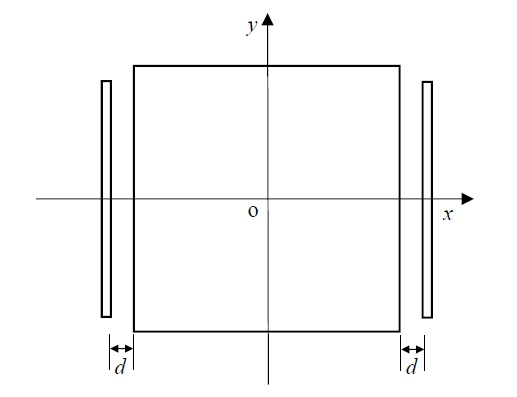}%
\caption{\label{sensor}  Schematic representation of the proof-mass with one electrode pair.}
\end{figure}

The generic model developed here is based on the gravitational reference sensor (GRS)~\cite{dolesi03} within the LISA Technology Package (LTP)~\cite{anza05} flown on LISA Pathfinder.  The basic sensor design~\cite{weber03} is a metallic cube surrounded by a number of electrode pairs enclosed within a metallic housing.  The electrode pairs perform different functions (signal injection, sensing and actuation in all six degrees of freedom) and are arranged symmetrically on either side of the proof-mass in all three axes.  One such pair is shown schematically in Figure~\ref{sensor}. 

In total the GRS has 9 electrode pairs providing only a partial coverage of the proof-mass surface area.  The electrodes are close-mounted into a metallic housing which then provides an effective ground plane coverage of the rest of the proof-mass. For the purposes of the electrostatic model the housing exposed directly to the proof-mass can be thought of as 3 additional grounded electrode pairs, one for each axis. The LTP used gold-coated cubic proof-masses with a side length 46\,mm.  The nominally symmetric  gaps, {\it d}, ranged from 2.9\,mm to 4.0\,mm depending on the function of the electrode pair.  Relevant to the charge sensitivity of this design will be the individual capacitances and, more importantly, the individual capacitance gradients. Finite element modelling is the most accurate way to find these with this type of complicated geometry~\cite{shaul04} but the design also lends itself to a simpler method of using parallel plate capacitor approximations and this gives results with better than 10\% accuracy which is sufficient for the setting of overall noise budgets. For example the total capacitance of the proof-mass with respect to its surroundings is 37\,pF using a parallel plate approximations (together with some allowance for fringing fields at the very edges because the housing is larger that the proof-mass) and 34.2\,pF using finite element methods~\cite{armano17}.  The total capacitance, $C_T$, is given by

\begin{equation}
\label{CT}
C_T = \sum\limits_{i=1}^{24}C_i   \hspace{2mm} {\rm where } \hspace{2mm} C_i=\frac{\epsilon A_i}{d_i} 
\end{equation}

where $\epsilon$ is the electrical permittivity within the gap, $A$ is the cross-sectional area, and values of $i$ from 1 to 18 refer to the electrodes proper and 19 to 24 refer to the housing ground planes.  The ground planes are designed to provide some degree of cross-coupling isolation and constitute some 61\% of the total capacitance. They also help to suppress fringing fields from around the edges of the individual electrodes coupling to the proof-mass making the parallel plate approximation  more accurate.

Figure~\ref{sensor} shows the proof-mass with a matched, and nominally identical, pair of electrodes along the $x-$axis.  Identifying these two electrodes from left to right with $i=1$ and $i=2$  the contribution to $C_T$ is simply $C_1 + C_2$.  The capacitance gradient along $x$, as experienced by the proof-mass moving along $x$, is 

\begin{equation}
\label{dcdx}
\frac{\partial C_T}{\partial x} = \frac{\partial C_1}{\partial x}+\frac{\partial C_2}{\partial x}=-\frac{C_1}{d_1}+\frac{C_2}{d_2}
\end{equation}
Therefore, if the two electrodes are identical and $d_1 = d_2=d$ and $C_1=C_2=C_x$, the gradient will be zero.  If the proof-mass is not at the centre, but is displaced by a small offset $x_o$ the capacitance gradient due to the matched pair of electrodes becomes

\begin{equation}
\label{dcdxwo}
\frac{\partial C_T}{\partial x} = \frac{C_x}{d} \frac{4x_o}{d}
\end{equation}

The first factor is the single-sided gradient and the second factor is $<<1$ ($\sim 0.01$ for the LTP experiment).  The total single-sided capacitance gradient in the most sensitive axis for the LTP design is $1.3$\,pF/m and comes from two pairs of sensing electrodes and the ground-plane coverage in that axis.  The two sensing electrodes contribute 20\% each.  If the two electrodes have an area mismatch, $\Delta A=A_2-A_1$, (through machining tolerances) then there will be an gradient even if the proof-mass is centred geometrically.  Equation~\ref{dcdxwo} becomes~\cite{sumner96}

\begin{equation}
\label{dcdxwoaa}
\frac{\partial C_T}{\partial x} = \frac{C_x}{d} \left( \frac{4x_o}{d}+\frac{\Delta A}{A} \right)
\end{equation}

In practice the co-ordinate system can be redefined to pass through the electrostatic centre to take account of any mismatch in the electrode (and housing) symmetry.  Hence equation~\ref{dcdxwo} will be used. 

In the absence of any free charge on the proof-mass the electrostatic force acting on the proof-mass along the axis $k$ will come from any voltages present on the electrodes according to:-

\begin{equation}
\label{zerochargef}
F_{k} = \frac{1}{2} \sum\limits_{i}\frac{\partial C_i}{\partial k} V_i^2
\end{equation}

There will be two types of deliberately applied voltage; the signal injection voltage at high frequency (needed to perform the position measurement) and somewhat lower frequency voltages to apply actuation forces to control the proof-mass in all six degrees of freedom.   The signal injection is applied in the two axes orthogonal to the sensitive direction.  The control voltages are applied on matched pairs of electrodes such that $\sum V_i =0$. 

The presence of a free charge, $Q$, on the proof-mass will result in two new electrostatic force terms to give an overall charge related force, $F_k^Q$

\begin{equation}
\label{echargef}
F_{Ek}^Q = \frac{Q^2}{2C_T^2}\frac{\partial C_T}{\partial k} - \frac{Q}{C_T}\sum\limits_{i} V_i\frac{\partial C_i}{\partial k}
\end{equation}

The first term is the interaction between $Q$ and its mirror charges induced in the electrodes.  The second term is the interaction between Q and {\it any} voltages present on the electrodes, both deliberately applied and stray~\cite{sumner04}. 

In addition to the electrostatic forces there will also be Lorentz forces due to the proof-mass (charge) motion through {\it any} magnetic fields present, both internal and external.  These will have the form

\begin{equation}
\label{lchargef}
F_{Lk}^Q = Q\left( \vec{v}_{SC} \times \vec{B}_{ext} + \vec{v}_{PM} \times \vec{B}_{ext} + \vec{v}_{PM} \times \vec{B}_{int}\right)_k
\end{equation}

where $\vec{v}_{SC}$ is the velocity of the spacecraft through the interplanetary magnetic field, $\vec{B}_{ext}$ , $\vec{v}_{PM}$ is the velocity of the proof-mass relative to the spacecraft and $\vec{B}_{int}$ is any magnetic field generated within, and locked to, the spacecraft.  As first pointed out by J-P Blaser~\cite{blaser00} the metallic enclosure around the proof-mass will act as a shield for it against the first term in equation~\ref{lchargef} through the generation of an effective Hall voltage across the enclosure.  For an ideal, completely closed and perfectly conducting enclosure the effect should be total.  However any apertures in the enclosure will result in magnetic field leakage and there will be some residual small efficiency, $\eta$, to be applied to the first term.  

\begin{equation}
\label{lchargefwe}
F_{Lk}^Q = Q\left( \eta \vec{v}_{SC} \times \vec{B}_{ext}  + \vec{v}_{PM} \times \left(\vec{B}_{ext}+\vec{B}_{int}\right) \right)_k
\end{equation}
  
A first estimate of the efficiency was evaluated at a very early LISA design phase~\cite{sumner97} giving $\eta<0.03$.  Since then the GRS design became that of a very much more enclosed proof-mass.

\section{\label{acc}Acceleration Noise from forces involving $Q$}

The forces arising due to a free-charge on the proof-mass are those in equations~\ref{echargef} and \ref{lchargefwe}. Any parameter within those force terms which exhibits a noisy behaviour will give rise to acceleration noise of the proof-mass.  Noise formulae for each term will be evaluated by considering which parameters in each one will exhibit noise~\cite{araujo03}.

\section{Electrostatic acceleration noise}

The first term in equation~\ref{echargef} can be combined with equation~\ref{dcdxwo} to give the acceleration as

\begin{equation}
\label{echargea1}
a_{Ek}^{Q_1} = \left(\frac{Q^2}{2MC_T^2}\frac{C_x}{d^2}4\right)x_o
\end{equation}

where $M$ is the mass of the proof-mass.  The acceleration is directly proportional to the displacement and the factor in parenthesis is thus an effective negative spring constant.  Note that $C_x$ and $d$ are nominal design values and are fixed.  $C_T$ on the other hand is the total capacitance and will depend on displacements in all three axes and angles through combinations of offsets in them~\cite{sumner97}.  However the dominant noise in $C_T$ resulting in acceleration noise along the $x$-axis  is directly from displacement noise of the proof-mass relative to the housing along the $x$-axis, charaterised by its amplitude spectral noise density, $S_x^{1/2}$.  This displacement noise also affect $x_o$ directly.  Finally the charge, $Q$, arises due to stochastic charge deposits from cosmic-rays and solar energetic particles~\cite{araujo05} and so there are three acceleration noise terms, $S_{a_j}^{1/2}$ arising out of equation~\ref{echargea1}.

From $S_x^{1/2}$, via $C_x$, there is
\begin{equation}
\label{sa1}
S_{a_1}^{1/2} = \frac{Q^2}{2MC_T^2}\left(\frac{C_x}{d^2}4\right)S_x^{1/2}
\end{equation}

From $S_x^{1/2}$, via $C_T$, there is
\begin{equation}
\label{sa2}
S_{a_2}^{1/2} = \frac{Q^2}{MC_T^3}\left(\frac{C_x}{d^2}4x_o\right)^2S_x^{1/2}
\end{equation}

From $S_Q^{1/2}$, via $Q$, there is
\begin{equation}
\label{sa3}
S_{a_3}^{1/2} = \frac{Q}{MC_T^2}\left(\frac{C_x}{d^2}4\right)x_oS_Q^{1/2}
\end{equation}
 
The second force term in equation~\ref{echargef} involves the interaction between $Q$ and voltages present on each electrode.  Due to the nominal symmetry of the sensor it is convenient to split this term into two parts.  The first due to common-mode voltages and the second due to differential mode voltages.

Forces due to common-mode voltages for a perfectly-centred proof-mass should sum up to zero as each electrode pair will combine with equal and opposite capacitance gradient.   However as already noted any displacement of the proof-mass from the electrostatic centre will prevent complete cancellation and for a common-mode voltage, $V_{c}$ there will be an acceleration 

\begin{equation}
\label{echargea2}
a_{Ek}^{Q_2} = \left(\frac{Q}{MC_T}\right)\left(\frac{V_{c}C_x'}{d^2}4x_o\right)
\end{equation}

where the result of the summation of the all relevant voltages and capacitances has been combined into the final factor and
$C_x'$ is the total capacitance from the electrodes proper; i.e. not including the ground plane.  This acceleration term involves $Q$, $C_T$, $V_c$ and $x_o$ all of which have noise associated with them.  Hence there are four acceleration noise components.

From $S_Q^{1/2}$ there is
\begin{equation}
\label{sa4}
S_{a_4}^{1/2} = \frac{1}{MC_T}\left(\frac{V_c C_x'}{d^2}4\right)x_oS_Q^{1/2}
\end{equation}

From $S_x^{1/2}$, via $C_T$, there is
\begin{equation}
\label{sa5}
S_{a_5}^{1/2} = \frac{Q}{MC_T^2}V_c\left(\frac{C_x}{C_x'}\right)\left(\frac{C_x'}{d^2}4x_o\right)^2S_x^{1/2}
\end{equation}
 
From $S_{V_c}^{1/2}$ there is
\begin{equation}
\label{sa6}
S_{a_6}^{1/2} = \frac{Q}{MC_T}\left(\frac{C_x'}{d^2}4x_o\right)S_{V_c}^{1/2}
\end{equation}
 
From $S_{x}^{1/2}$ there is
\begin{equation}
\label{sa7}
S_{a_7}^{1/2} = \frac{QV_c}{MC_T}\left(\frac{C_x'}{d^2}4\right)S_{x}^{1/2}
\end{equation}

Forces due to differential-mode voltages for a perfectly-centred proof-mass will not sum up to zero.  Instead they will act through the individual capacitance gradient(s) of whichever electrode(s) they are present on, including the ground planes of the housing.   Differential mode voltages could arise either through mismatch in applied voltages otherwise intended to balance (due to scale factors or incoherent noise), or through uncontrolled stray potentials associated with individual surfaces~\cite{speake96, carbone03, weber07, pollack08, antonucci12, yin14}. For simplicity it will be assumed here that only one surface, labelled simply as $i$, is implicated with a stray voltage, $V_i$.  Then 

\begin{equation}
\label{echargea3}
a_{Ek}^{Q_3} = \left(\frac{Q}{MC_T}\right)V_i\frac{\partial C_i}{\partial k}
\end{equation}

This acceleration term involves $Q$, $C_T$, $\partial C_i/\partial k$ and $V_i$ all of which have noise associated with them.   Hence there are four more acceleration noise components.

From $S_Q^{1/2}$ there is
\begin{equation}
\label{sa8}
S_{a_8}^{1/2} = \frac{1}{MC_T}V_i\frac{\partial C_i}{\partial k}S_Q^{1/2}
\end{equation}
 
From $S_x^{1/2}$, via $C_T$, there is
\begin{equation}
\label{sa9}
S_{a_9}^{1/2} = \frac{Q}{MC_T^2}V_i\frac{\partial C_i}{\partial k}\left(\frac{C_x}{d^2}4x_o\right)S_x^{1/2}
\end{equation}

From $S_x^{1/2}$, via $\partial C_i/ \partial k$, there is
\begin{equation}
\label{sa10}
S_{a_{10}}^{1/2} = \frac{Q}{MC_T}V_i\frac{\partial^2 C_i}{\partial k^2}S_x^{1/2}
\end{equation}

From $S_{V_i}^{1/2}$ there is
\begin{equation}
\label{sa11}
S_{a_{11}}^{1/2} = \frac{Q}{MC_T}\frac{\partial C_i}{\partial k}S_{V_i}^{1/2}
\end{equation}

\subsection{Lorentz force acceleration noise}

From equation~\ref{lchargefwe} the maximum acceleration experienced by the charged proof-mass due to magnetic field interactions is 

\begin{equation}
\label{lchargefwea}
a_{Lk}^Q = \frac{Q}{M}\left( \eta v_{SC} B_{ext}  + v_{PM} \left(B_{ext}+B_{int}\right)\right)_k
\end{equation}
  
This acceleration involves $Q$, $v_{SC}$, $ B_{ext}$, $v_{PM}$ and $ B_{int}$ all of which have noise associated with them.   Indeed even $\eta$ could have a time dependent, and hence noisy behaviour, due to pitch angle variation in the external B field, but this will not be addressed here.  Hence there are five more acceleration noise components.

From $S_Q^{1/2}$ there is
\begin{equation}
\label{sa12}
S_{a_{12}}^{1/2} = \frac{1}{M}\left( \eta v_{SC} B_{ext}  + v_{PM} \left(B_{ext}+B_{int}\right)\right)S_{Q}^{1/2}
\end{equation}
 
From $S_{v_{SC}}^{1/2}$ there is
\begin{equation}
\label{sa13}
S_{a_{13}}^{1/2} = \frac{Q}{M}\left( \eta B_{ext} \right)S_{v_{SC}}^{1/2}
\end{equation}
 
From $S_{B_{ext}}^{1/2}$ there is
\begin{equation}
\label{sa14}
S_{a_{14}}^{1/2} = \frac{Q}{M}\left( \eta v_{SC}  + v_{PM}\right)S_{B_{ext}}^{1/2}
\end{equation}
 
From $S_{v_{PM}}^{1/2}$ there is
\begin{equation}
\label{sa15}
S_{a_{15}}^{1/2} = \frac{Q}{M}\left( B_{ext}+B_{int}\right)S_{v_{PM}}^{1/2}
\end{equation}
 
From $S_{B_{int}}^{1/2}$ there is
\begin{equation}
\label{sa16}
S_{a_{16}}^{1/2} = \frac{Q}{M}\left( v_{PM}\right)S_{B_{int}}^{1/2}
\end{equation}

There are thus sixteen individual contributions to the overall charge-induced acceleration noise and these will be  in the next section to find the overall noise budget.

\section{\label{budget}Overall Acceleration Noise Budget Evaluation}  

The linear acceleration noise will be evaluated along the sensitive $x$-direction for each of the contributions.

\subsection{Amplitude spectral noise distributions within electrostatic noise contributions}

In the sensitive direction the nominal amplitude spectral density of the position noise to be used in  equations~\ref{sa1},~\ref{sa2},~\ref{sa5},~\ref{sa7},~\ref{sa9} and \ref{sa10} is taken directly from the requirements detailed in the LISA proposal~\cite{Amaro17} for the interferometric ranging, assuming equal contributions from the two proof-masses in a link.

\begin{equation}
\label{sxdisp}
S_x^{1/2} \leq 7.1 \times 10^{-12}\sqrt{1+\left(\frac{2\times 10^{-3}}{f}\right)^4} \hspace{2mm} \rm{m/\sqrt{Hz}}
\end{equation}

where $f$ is the frequency. However, this is a requirement on the differential noise between two widely separated proof-masses forming a link, and it does not directly give the relative spatial noise behaviour of one proof-mass within its own spacecraft enclosure which will depend on the inherent platform stability, the local displacement measurement and the closed-loop drag-free performance used to control the spacecraft motion.  This has been studied in detail for LISA Pathfinder~\cite{armano18e} and we use results from that study to provide representative performance data.  Figure 2 in \cite{armano18e} shows an amplitude spectral density between 0.05mHz and 30mHz which we approximate to 

\begin{equation}
\label{sxdisplpf}
S_x^{1/2} \approx 3 \times 10^{-5}f^2 \hspace{2mm} \rm{m/\sqrt{Hz}}
\end{equation}

Although equations~\ref{sxdisp} and \ref{sxdisplpf} have very different forms their values at 1mHz are almost identical.
In the less-sensitive directions, relevant to equation~\ref{sa15},  the displacement noise requirement is relaxed to 5\,nm/$\rm{\sqrt{Hz}}$ and the LISA Pathfinder data (Figure 5 in \cite{armano18e}) show an in-flight performance of  $S_{y,z}^{1/2}$ in the range  $ 0.3 - 2 \times 10^{-8} \rm{m/\sqrt{Hz}}$ between .01 and 30\,mHz \footnote{Using the `State reproduction from the model' as a true indicator of this noise}.

The charge noise, $S_Q^{1/2}$ to be used in equations~\ref{sa3},~\ref{sa4},~\ref{sa8} and \ref{sa12} was first derived using Monte-Carlo simulations~\cite{araujo05} but has since been modified by in-orbit measurements made by LISA Pathfinder~\cite{armano17}.  The charging process is due to the interaction of cosmic-rays within the proof-mass and surrounding structures.  The cosmic-rays have random arrival times and each cosmic-ray charges the proof-mass by an amount drawn at random from a broad distribution~\cite{araujo05}. Hence the charging rate will exhibit shot noise, but at a level exceeding that expected for single charge contributions; i.e. $S_I^{1/2}=e\sqrt{2\lambda_{eff}}$ where $\lambda_{eff} >> \dot{Q}/e$ is the effective single-charge noise rate.  The average charging rate seen on the two proof-masses in LISA Pathfinder was $+23.7$\,charges/s with a mean effective noise rate $\lambda_{eff}=1100$\,charges/s.  The charge, $Q$ is then the result of integrating the current.  Consequently the charge spectral noise density relevant to equations~\ref{sa3}, \ref{sa4}, \ref{sa8} and \ref{sa12} is: 

\begin{equation}
\label{sq}
S_Q^{1/2}= \frac{1}{2\pi f}S_I^{1/2}=\frac{e}{2\pi f} \sqrt{2 \lambda_{eff}}\hspace{2mm} \rm{C/\sqrt{Hz}}
\end{equation}

This charge spectral noise density is relevant to equations~\ref{sa3},~\ref{sa4},~\ref{sa8} and \ref{sa12}. 

The appropriate form for the amplitude spectral density of the common-mode voltage, $S_{V_c}^{1/2}$, depends on how the voltages are produced and applied.  The lowest noise performance, at least from the point of view of common-mode noise, will be if the noise on each of the applied voltages is independent. The common mode sum will then be roughly a factor $\sqrt{n}$ higher whereas if the noise sources are correlated (i.e derived from a common reference) the overall noise scales as $n$. Typically high-precision voltage references have noise at the ppm level, which will be assumed for equation~\ref{sa6}.

In equation~\ref{sa11} the spectral noise density for differential mode (stray) voltages is required.  Stray voltages on gold surfaces have been measured up to 100\,mV levels.   Early measurements of the noise associated with such stray voltages gave a white spectral density of $30\mu$V$/\rm{\sqrt{Hz}}$ above 0.1\,mHz but rising to lower frequencies~\cite{pollack08}. More recently LISA Pathfinder has made a new more representative measurement in space, which included both stray voltages and low-frequency applied voltage noise~\cite{armano17}.  The result was no longer white above 0.1mHz but could be well fitted by a form

\begin{equation}
\label{svi}
S_{V_i}^{1/2} =\sqrt{ \frac{8.83\times 10^{-16}}{f^2} + \frac{2.73 \times 10^{-13}}{f}}   \hspace{2mm} \rm{V/\sqrt{Hz}}
\end{equation}

It should be noted that this mathematically assigns the effect of all the differential stray voltages to just one of the two sensing electrodes along the $x$-direction, following the convention adopted in ~\cite{antonucci12}.  If this assumption is relaxed and the stray voltages, for example, are uniformly distributed between both sensing electrodes and the ground plane the overall result would be the same as obtained with equation~\ref{svi} as the voltage fluctuations would be reduced in proportion to the electrode areas and the gaps are the same for each.

\subsection{Amplitude spectral distributions within Lorentz force noise contributions}

The spacecraft velocity noise within equation~\ref{sa13} will be due to solar wind forces.  To generate an acceleration noise along the $x$-axis the spacecraft velocity noise must be in a perpendicular direction. In that direction drag-free satellite control will suppress the actual movement of the spacecraft using the external micro-thrusters using measurements from capacitance sensing of the proof-mass.  The resulting residual motion of the spacecraft relative to the local gravitational frame will be from a combination of uncertainty (noise) in the capacitance measurement and noise in the thrusters giving imperfect drag-free control.  The requirement specification for measurement of relative proof-mass displacement within the spacecraft calls for 5nm/$\rm{\sqrt{Hz}}$. The in-flight LISA Pathfinder platform stability performance, including the closed-loop drag-free, is shown in Figure 6 of \cite{armano18e}.  The spacecraft acceleration curves recovered from the simulation have a characteristic 'V' shape which, for the relevant transverse axes, can be approximated to within a factor of two, to a functional form

\begin{equation}
\label{sasc}
S_{a_{SC}}^{1/2}=\frac{3 \times 10^{-18}}{f} + {2 \times 10^{-7}}f^2 \hspace{5mm} \rm{ms^{-2}/\sqrt{Hz}}
\end{equation}

From this the amplitude spectral density of the velocity noise is 

\begin{equation}
\label{svsc}
S_{v_{SC}}^{1/2}=\frac{3 \times 10^{-18}}{2 \pi f^2} + \frac{{2 \times 10^{-7}}f}{\pi} \hspace{5mm} \rm{ms^{-1}/\sqrt{Hz}}
\end{equation}

The power spectral density in the external interplanetary magnetic field at 1AU can be found in \cite{tu95} and \cite{forsyth96}.  Larger fluctuations are seen for the magnetic field components than for the total field as they are anti-correlated.  A $1/f^n$ fit to the noisiest component in their data over the relevant frequency range gives 

\begin{equation}
\label{sbext}
S_{B_{ext}}^{1/2} =\frac{0.17 \times 10^{-9}}{f^{0.8}} \hspace{5mm} \rm{T/\sqrt{Hz}} 
\end{equation}

Equation~\ref{sa15} requires the velocity noise spectrum for the proof-mass relative to its housing. It does not matter whether the proof-mass is actually moving but rather that there is relative motion between the proof-mass and the magnetic field source.  The requirement specification for measurement of relative proof-mass displacement within the spacecraft calls for 5nm/$\rm{\sqrt{Hz}}$. The in-flight LISA Pathfinder platform stability performance, including the closed-loop drag-free, gave a resultant maximum amplitude spectral density of 20nm/$\rm{\sqrt{Hz}}$.  Conservatively treating this as frequency independent implies,

\begin{equation}
\label{svpm}
S_{v_{PM}}^{1/2}=4 \pi f \times 10^{-8} \hspace{5mm} \rm{ms^{-1}/\sqrt{Hz}}
\end{equation}

The magnetic field noise was measured in situ by magnetometers on board LISA Pathfinder. At low frequencies ($<$ 2 mHz) the measurements merged into the interplanetary magnetic field data~\cite{hollington18}.  At higher frequencies the amplitude spectral density was white and close to the pre-flight requirement on magnetic field noise produced at the proof-mass by local instrumentation, of $<5\rm{nT/\sqrt{Hz}}$~\cite{mateos09}.  The magnetic field noise measured by LISA Pathfinder would have included spatial variations in the field converted into time dependent variations due to the spacecraft motion. These should be fairly representative of that expected for LISA given the L1 orbit around the Sun is close to 1AU.

\subsection{Relevant GRS Parameters}

Table~\ref{Parameters} lists the nominal values of all the parameters used in estimating the acceleration noise from each of the 16 terms identified.

\begin{table}[h]
\begin{center}
\caption{\label{Parameters}Relevant GRS parameters used for charge-induced noise evaluation}
\begin{tabular}{cccc}
\hline\hline
Parameter & Value & Units & Equations \\ \hline
M &  1.928 & kg & [\ref{sa1} -- \ref{sa16}]\\
$C_T$ & 36.9 & pF &  [\ref{sa1} -- \ref{sa11}]\\
$C_x$ & 5.36 & pF & [\ref{sa1} -- \ref{sa7}],~\ref{sa9}\\
$d$ & 4.0 & mm & [\ref{sa1} -- \ref{sa7}],~\ref{sa9}\\
$x_o$ & 10 & $\rm{\mu}$m & [\ref{sa2} -- \ref{sa6}],~\ref{sa9}\\
$V_c$ & 1 & V & \ref{sa4}, \ref{sa7} \\
$V_i$ & 3$^a$ & mV & [\ref{sa8} -- \ref{sa10}] \\
$\partial C_i / \partial k$ & 540$^b$  & pF/m & \ref{sa8}, \ref{sa9}, \ref{sa11} \\
$\partial^2 C_i / \partial k^2$ & 135,000$^c$ & pF/m$^2$ & \ref{sa10}  \\
$ \eta$ & .01$^d$   & - & [\ref{sa12} -- \ref{sa14}]  \\
$v_{SC}$ & $3 \times 10^4$ & m/s & \ref{sa12}, \ref{sa14}  \\
$B_{ext}$ & 3$^e$ & nT &  \ref{sa12}, \ref{sa13}, \ref{sa15} \\
$v_{PM} $ &  $10\pi f^{(3/2)} $$^f$  & nm/s & \ref{sa12}, \ref{sa14}, \ref{sa16} \\
$B_{int} $ & 10$^g$ \footnote{See ~\cite{mateos09}} & $\rm{\mu}$T & \ref{sa12}, \ref{sa15} \\
\hline\hline \\
\end{tabular}
\end{center}
$^a$ Assuming dc compensation following~\cite{weber07} as demonstrated on LISA Pathfinder~\cite{armano17} \\
$^b$ Expressed as an effective  single-sided gradient from the two sensing electrodes as assumed for LISA Pathfinder~\cite{armano17} \\
$^c$ $\left(\partial C_i / \partial k \right) /d$ \\
$^d$ Taking the {\it very} conservative early estimate~\cite{sumner97} reduced by a further factor of 3 for the newer more closed designs \\ 
$^e$ See~\cite{tu95} \\
$^f$ Estimated from $S_{y,z}^{1/2}$. Note $f$ in Hz \\
$^g$ See~\cite{mateos09} \\
\end{table}

\subsection{Overall noise performance}
\label{performance}
   
\begin{figure}[h]
\begin{center}
\includegraphics[width=0.8\textwidth]{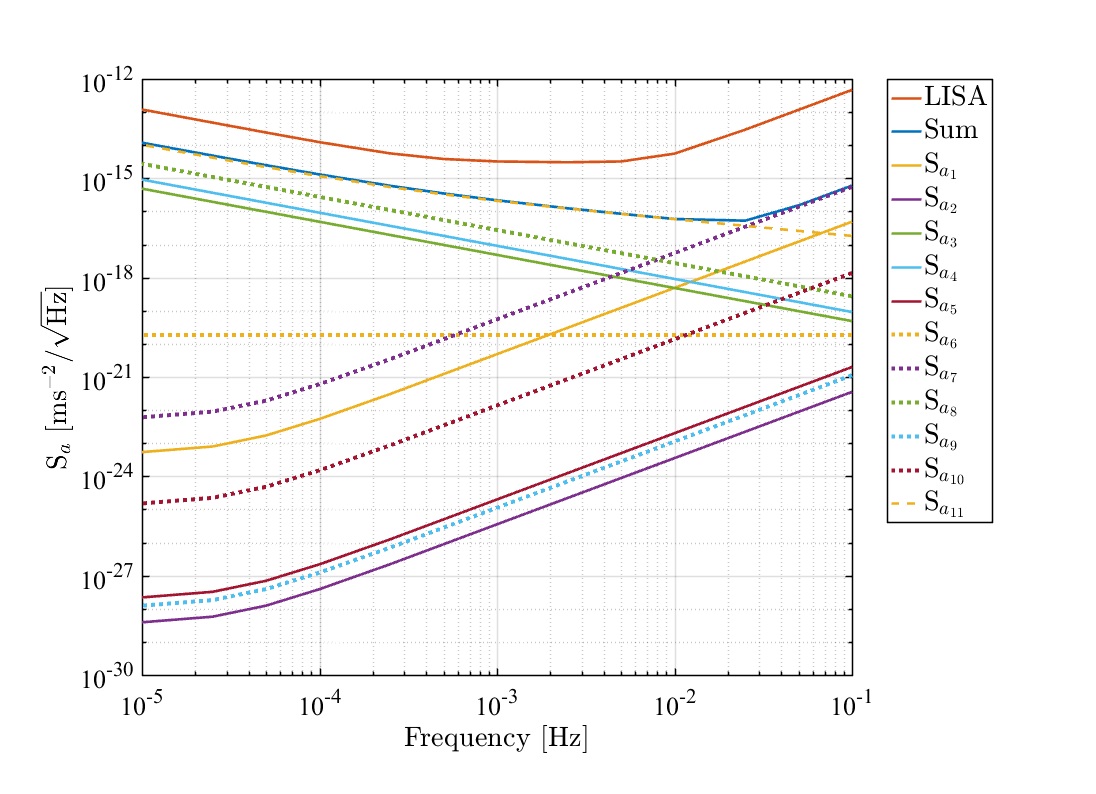}%
\caption{\label{enoise}  Acceleration noise density associated with electrostatic contributions, $S_{a1}^{1/2}$ to $S_{a11}^{1/2}$, in continuous discharge mode assuming $Q = 5 \times 10^6$ charges.}
\end{center}
\end{figure}

Figure~\ref{enoise} shows the contributions from the electrostatic terms, $S_{a1}^{1/2}$ through $S_{a11}^{1/2}$, to the acceleration noise spectral density.  Also shown is the performance requirement for the overall LISA GRS acceleration noise for each proof-mass~\cite{Amaro17}. 

\begin{equation}
\label{sabudget}
S_a^{1/2} \leq 3 \times 10^{-15}\sqrt{1+\left(\frac{0.4}{f}\right)^2}\sqrt{1+\left(\frac{f}{8}\right)^4} \rm{ms^{-2}/\sqrt{Hz}}
\end{equation}

The `Sum' curve is the quadrature sum of all terms. In principle terms driven by the same noisy parameter should be added linearly but in practice they also contain factors whose absolute sign is uncertain.  The values of all the parameters have been set to those appropriate for continuous discharge during science mode operations, as shown in table~\ref{Parameters}, with a charge level, $Q$, set to $5\times 10^6$ elementary charges.

$S_{a11}^{1/2}$ is the most dominant term, and is almost coincident with the `Sum' curve on the plot.  It arises due to the interplay between $Q$ and the spectral noise density from the differential (stray+ applied) voltages.
$S_{a8}^{1/2}$ and $S_{a4}^{1/2}$ are next most dominant terms, and on the plot there are coincidentally almost equal. $S_{a8}^{1/2}$ comes from the interaction between differential (stray) voltages and the charge noise spectral density.  $S_{a4}^{1/2}$ comes from the interaction between common mode (injection) voltages and the charge noise spectral density. The total spectral noise density from all electrostatic terms is comfortably below the performance requirements, partly, it should be noted, due to the stray voltage compensation scheme~\cite{weber07}, now successfully demonstrated in flight with LISA Pathfinder~\cite{armano17}.

\begin{figure}[h]
\begin{center}
\includegraphics[width=0.8\textwidth]{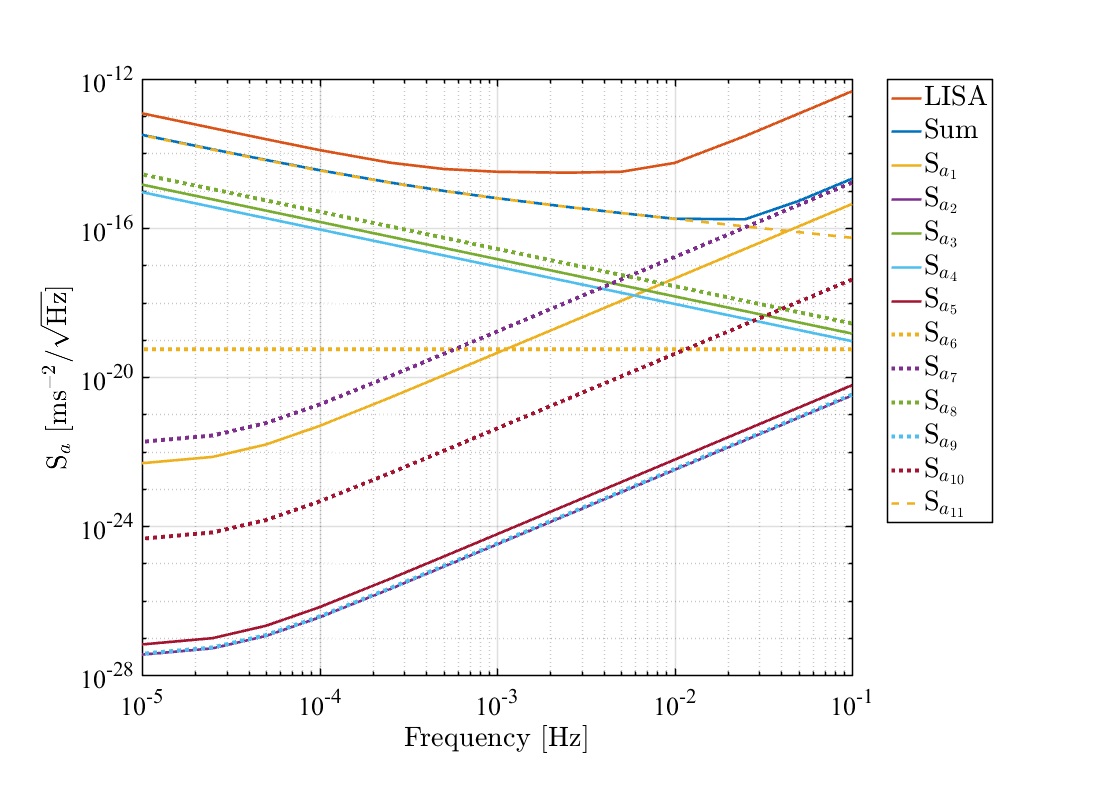}%
\caption{\label{enoise2}  Maximum acceleration noise density associated with electrostatic contributions, $S_{a1}^{1/2}$ to $S_{a11}^{1/2}$, during science operations using the intermittent rapid discharge mode with $Q_{max} = 1.5 \times 10^7$ charges.}
\end{center}
\end{figure}

Figure~\ref{enoise2} shows a similar plot but with $Q=1.5 \times 10^7$ elementary charges, which is the maximum charge expected in science mode if an intermittent discharge scheme is adopted.  $S_{a11}^{1/2}$ has increased proportionately to $Q$.  The `Sum' remains below the requirement specification of the GRS. The electrostatic noise approaches the requirement specification closest at low frequencies, below a mHz. 

\begin{figure}[h]
\begin{center}
\includegraphics[width=0.8\textwidth]{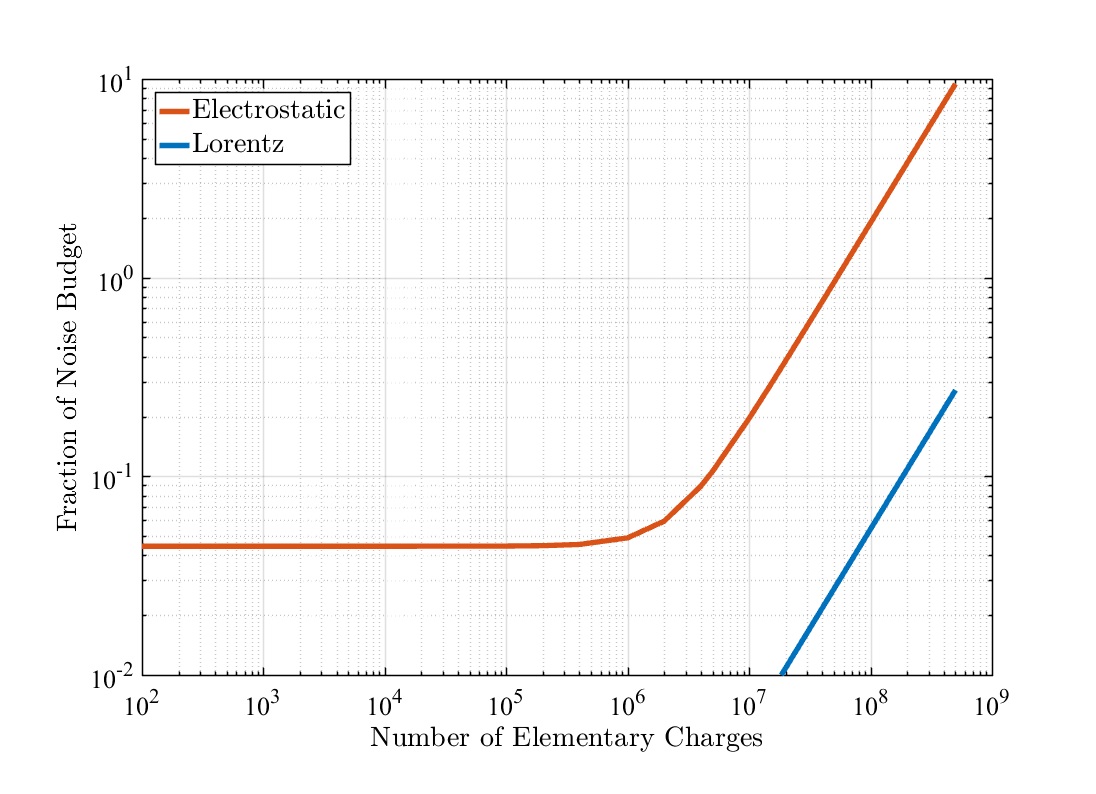}%
\caption{\label{ratios}  Fraction of the LISA GRS acceleration noise budget absorbed by the charge induced noise as a function of $Q$ at 0.1\,mHz.}
\end{center}
\end{figure}

 Figure~\ref{ratios} shows the fraction of the GRS acceleration noise budget absorbed by the charge induced noise at 0.1\,mHz as a function of $Q$. Below $Q=10^6$\, charges the fraction becomes charge independent, but non-zero.  This is due to $S_{a8}^{1/2}$ and $S_{a4}^{1/2}$ which do not depend on absolute charge, but do depend on the charge noise,  and between them contribute $\sim 3 \times 10^{-16}\,\rm{ms^{-2}/\sqrt{Hz}}$ at 0.1mHz.  $Q$ can increase up to $5 \times 10^7$ charges before the GRS budget is fully absorbed by charge induced noise.  Also shown in figure~\ref{ratios} is the overall contribution from the Lorentz force noise terms, $S_{a12}^{1/2}$ through $S_{a16}^{1/2}$. It can be seen that these are insignificant compared to the electrostatic terms, although the dominant term within the Lorentz noise benefits from the (fortuitous) shielding factor, $\eta$ and, without that, the Lorentz force noise would dominate above $Q\sim 10^6$ charges.

\begin{figure}[h]
\begin{center}
\includegraphics[width=0.8\textwidth]{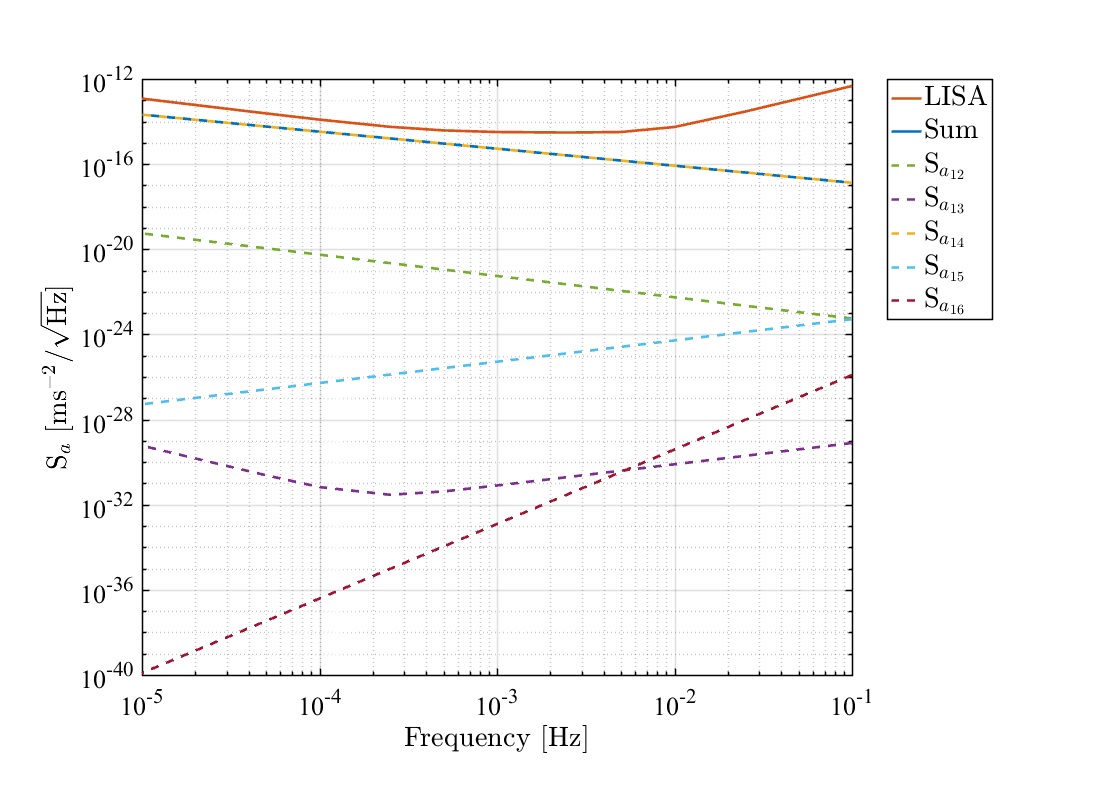}%
\caption{\label{lnoise}   Acceleration noise density associated with Lorentz contributions, $S_{a12}^{1/2}$ to $S_{a16}^{1/2}$.  $Q = 5 \times 10^8$ is used to enhance these terms.}
\end{center}
\end{figure}

Figure~\ref{lnoise} shows the individual Lorentz force noise components with a relatively high $Q$ of  $5 \times 10^8$ charges.  The dominant term is $S_{a14}^{1/2}$ which is coincident with the 'Sum' curve and is the interaction between $Q$ and the external magnetic field fluctuations, after allowing for an assumed  99\% effective shielding factor ($\eta = 0.01$) from the Hall effect in the metallic housing.

\section{\label{discussion}Discussion}
A comprehensive evaluation of the linear acceleration spectral noise density terms arising from free charge residing on the proof-mass has been carried out.  Sixteen first-order terms have been identified including both electrostatic interactions and magnetic field interactions.  All terms have been informed by the recent in flight experience with LISA Pathfinder, during which some of the experiments were specifically designed to help consolidate the charge induced noise understanding.

\subsection{Acceleration noise}
The dominant charge related terms contributing to the GRS acceleration noise are electrostatic terms. 
The most dominant involves the differential voltage fluctuations in equation~\ref{svi} while the next two involve the noisy charging process via equation~\ref{sq}.  

For both continuous discharging and intermittent discharging modes of operation the charge related noise remains within the overall budget, for the parameters adopted in table~\ref{Parameters}.  Most of the key parameters in that table are based on LISA Pathfinder which was ultimately able to show compliance with the LISA budget~\cite{armano18d}.

From figure~\ref{ratios} it can be seen that there is no advantage to reducing the charge much below $10^6$ charges from the point of view of acceleration noise.  Hence the continuous discharge mode only needs to ensure $Q < 5 \times 10^6$ charges. 

\subsection{Other charge related effects}
\subsubsection{Forces}
Forces will build up as charge is accumulating on the proof-mass.   The size of these forces could in principle cause bias effects.  The three basic forces are given in table~\ref{forcess} for various values of $Q$.  Note there is a coincidental equality between the sizes of the second two forces given the parameter values used in Table~\ref{Parameters}

\begin{table}[h]
\begin{center}
\caption{\label{forcess}Charge induced forces}
\begin{tabular}{cccc}
\hline \\
$Force$ & Q & Value & Units \\ \hline
$\frac{Q^2}{2C_T^2}\frac{\partial C_T}{\partial k}  $ &  0 & 0 & N\\
$$ & 1E6 & $1.26 \times 10^{-16}$ &  N \\
$$ & 1.5E7 & $2.83 \times 10^{-14}$ & N \\
$$ & 5E8 & $ 3.15 \times 10^{-11}$ & N \\ \hline
$\frac{Q}{C_T}\sum V_i \frac{\partial C_i}{\partial k}  $ &  0 & 0 & N\\
$$ & 1E6 & $7.10 \times 10^{-15}$ &  N \\
$$ & 1.5E7 & $1.07 \times 10^{-13}$ & N \\
$$ & 5E8 & $ 3.55 \times 10^{-12}$ & N \\ \hline
$\frac{Q}{C_T}V_{com}\frac{4C_k}{d^2} \delta x  $ &  0 & 0 & N\\
$$ & 1E6 & $7.10 \times 10^{-15}$ &  N \\
$$ & 1.5E7 & $1.07 \times 10^{-13}$ & N \\
$$ & 5E8 & $ 3.55 \times 10^{-12}$ & N \\ \hline

\end{tabular}
\end{center}
\end{table}

\subsubsection{Spring constant}
The presence of free charge on the proof-mass will introduce new terms into the spring constant and these must be small enough not to unduly affect the dynamics of the system.  Typical spring constants observed on LISA Pathfinder were $\sim 10^{-6}$\,N/m ~\cite{armano16} and from Table~\ref{springs} it can be seen that charge induced electrostatic spring terms will not become problematic until $Q$ approaches $10^8$ charges.

\begin{table}[h]
\caption{\label{springs}Charge induced spring constants}
\begin{center}
\begin{tabular}{cccc}
\hline \\
$Force / (Spring Constant)$ & Q & Value & Units \\ \hline
$\frac{Q^2}{2C_T^2}\frac{\partial C_T}{\partial k}  $ &  0 & 0 & N/m\\
$$ & 1E6 & $1.26 \times 10^{-11}$ &  N/m \\
$\left( \frac{1}{2}\frac{Q^2}{C_T^2}\frac{4C_x}{d^2}+\frac{Q^2}{C_T^3}\left( \frac{4C_x}{d^2} \delta x \right)^2 \right) $ & 1.5E7 & $2.83 \times 10^{-9}$ & N/m \\
$$ & 5E8 & $ 3.15 \times 10^{-6}$ & N/m \\ \hline
$\frac{Q}{C_T}\sum V_i \frac{\partial C_i}{\partial k} $ &  0 & 0 & N/m\\
$$ & 1E6 & $1.78 \times 10^{-12}$ &  N/m \\
$\left( \frac{QV_i}{MC_T^2}\left(\frac{C_x}{d^2}4\delta x+\frac{\partial^2 C_i}{\partial x^2}\right)\right)$ & 1.5E7 & $2.66 \times 10^{-11}$ & N/m \\
$$ & 5E8 & $ 8.88 \times 10^{-10}$ & N/m \\ \hline
$\frac{Q}{C_T}V_{com}\frac{4C_k}{d^2} \delta x $ &  0 & 0 & N/m\\
$$ & 1E6 & $7.10 \times 10^{-10}$ &  N/m \\
$\left( \frac{Q}{C_T}V_{com}\frac{4C_x}{d^2}+\frac{Q}{C_T^2}V_{com} \left( \frac{4C_x}{d^2} \delta x \right)^2 \right) $ & 1.5E7 & $1.07 \times 10^{-8}$ & N/m \\
$$ & 5E8 & $ 3.55 \times 10^{-7}$ & N/m \\ \hline

\end{tabular}
\end{center}
\end{table}

\subsubsection{Data artefacts}
During solar-quiet times the charging rate will follow the ambient cosmic-ray environment~\cite{armano18b}.  The cosmic-ray rate will be fairly steady but with quasi-periodic modulations at the few percent level with the occasional larger short-term Forbush depressions~\cite{armano18c}. The overall trend in the charge on the proof-mass will be linear with time if the charge is allowed to build-up.  This linear charge build-up will give rise to forces which grow both linearly and quadratically with time and this will be happening quasi-independently on the six proof-masses within the LISA constellation.  In the case of an intermittent discharge scheme, with a periodic discharge sequence, this could produce fourier components in the data which could become problematic~\cite{shaul05}.

\ack
TJS acknowledges support from the Leverhulme Trust (EM-2019-070\textbackslash 4).

\end{document}